\documentclass{INTERSPEECH2023}

\interspeechcameraready

\usepackage{tikz}
\usepackage{amssymb}
\usepackage{algorithm,setspace}
\usepackage[noend]{algpseudocode}
\usepackage{booktabs}
\usepackage{pifont}
\usepackage{multirow}
\usepackage{tabularx,ragged2e}
\newcolumntype{Y}{>{\centering\arraybackslash}X}

\usepackage{pgfplots}
    \usetikzlibrary{calc,positioning,patterns}
    \pgfplotsset{compat=1.3}
\usepackage{pgfplotstable}
\usepackage[nolist, nohyperlinks]{acronym}

\newcommand{\mf}[1]{\bar{\mathbf{#1}}}
\newcommand{\spec}[1]{\mathbf{#1}}
\newcommand{\tim}[1]{{#1}}

\newcommand{\ssp}[0]{\text{}}

\tikzstyle{block} = [draw, fill=blue!14, rectangle, minimum height=3em, minimum width=5em, align=center]
\tikzstyle{sum} = [draw, fill=white, circle, node distance=1cm]
\tikzstyle{input} = [coordinate]
\tikzstyle{output} = [coordinate]
\tikzstyle{pinstyle} = [pin edge={to-,thin,black}]
\tikzstyle{branch}=[fill,shape=circle,minimum size=3pt,inner sep=0pt]
\tikzstyle{connarrow}=[-latex, line width=1pt]
\tikzstyle{connline}=[-, line width=1pt]

\title{Extending DNN-based Multiplicative Masking to Deep Subband Filtering for Improved Dereverberation}
\name
 {Jean-Marie Lemercier, Julian Tobergte, Timo Gerkmann\thanks{This work has been funded by the Federal Ministry for Economic Affairs and Climate Action, project 01MK20012S, AP380. The authors are responsible for the content of this paper.}}
	\address{Signal Processing (SP), Universität Hamburg, Germany
	\email{\tt \small \{firstname.lastname\}@uni-hamburg.de}
	}
	
\begin{document}

\begin{acronym}
\acro{dsfe}[DSFE]{deep subband filtering extension}
\acro{stft}[STFT]{short-time Fourier transform}
\acro{istft}[iSTFT]{inverse short-time Fourier transform}
\acro{dnn}[DNN]{deep neural network}
\acro{pesq}[PESQ]{Perceptual Evaluation of Speech Quality}
\acro{wpe}[WPE]{weighted prediction error}
\acro{psd}[PSD]{power spectral density}
\acro{rir}[RIR]{room impulse response}
\acro{snr}[SNR]{signal-to-noise ratio}
\acro{si}[SI-]{scale-invariant}
\acro{sir}[SIR]{signal-to-interference ratio}
\acro{sdr}[SDR]{signal-to-distortion ratio}
\acro{sar}[SAR]{signal-to-artifacts ratio}
\acro{lstm}[LSTM]{long short-term memory}
\acro{polqa}[POLQA]{Perceptual Objectve Listening Quality Analysis}
\acro{sdr}[SDR]{signal-to-distortion ratio}
\acro{estoi}[ESTOI]{Extended Short-Term Objective Intelligibility}
\acro{elr}[ELR]{early-to-late reverberation ratio}
\acro{tcn}[TCN]{temporal convolutional network}
\acro{rls}[RLS]{recursive least squares}
\acro{asr}[ASR]{Automatic speech recognition}
\acro{ha}[HA]{hearing aid}
\acro{ci}[CI]{cochlear implant}
\acro{mac}[MAC]{multiply-and-accumulate}
\acro{drr}[DRR]{direct-to-reverberation ratio}
\acro{tf}[TF]{time-frequency}
\end{acronym}

\maketitle

\begin{abstract}
In this paper, we present a scheme for extending deep neural network-based multiplicative maskers to deep subband filters for speech restoration in the time-frequency domain. 
The resulting method can be generically applied to any deep neural network providing masks in the time-frequency domain, while requiring only few more trainable parameters and a computational overhead that is negligible for state-of-the-art neural networks.
We demonstrate that the resulting deep subband filtering scheme outperforms multiplicative masking for dereverberation, while leaving the denoising performance virtually the same.
We argue that this is because deep subband filtering in the time-frequency domain fits the subband approximation often assumed in the dereverberation literature, whereas multiplicative masking corresponds to the narrowband approximation generally employed for denoising. 
\end{abstract}


\section{Introduction}
\label{sec:intro}
In modern communication devices, recorded speech is corrupted when clean speech sources are affected by interfering speakers, background noise and room acoustics.
Speech restoration aims to recover clean speech from the corrupted signal, whereby two distinct tasks, denoising and dereverberation, are considered here \cite{Naylor2011, Godsill1998DigitalAR}.

Traditional speech restoration algorithms are based on statistical methods, exploiting properties of the target and interfering signals to discriminate between them~\cite{gerkmann2018book_chapter}. These include linear prediction \cite{Nakatani2008b}, 
spectral enhancement \cite{Habets2007}, inverse filtering \cite{Kodrasi2014}, and cepstral processing \cite{Gerkmann2011}.
Modern approaches rely mostly on machine learning. In this field, predictive methods, learning a one-to-one mapping between corrupted and clean speech through a \ac{dnn}, are most popular \cite{MurphyBook2, Wang2018SUpervised}.
A large portion of \acp{dnn} used in speech restoration are trained for mask estimation, i.e. they learn a mask value to be applied to each single bin of the signal, either in a learnt domain \cite{Luo2019ConvTasNet} or in the \ac{tf} domain \cite{Li2022Gagnet, Williamson2017j}. On the opposite, some approaches employ deep filtering \cite{Mack2020DeepFiltering}, which means that their final stage involves a convolution between the input signal and a learnt multi-frame \ac{tf} filter \cite{Schroeter2022DeepFilterNet, Schroeter2020CLCNet, Lv2020DCCRNPlus, 
Tammen2021DeepMFMVDR, Heymann2018, Kinoshita2017, Lemercier2022}. In \cite{Tammen2021DeepMFMVDR}, this filter is parameterized as a multi-frame MVDR \cite{Huang2012MFMVDR} for denoising. A DNN-parameterized weighted prediction error subband filter is proposed in \cite{Kinoshita2017, Heymann2018, Lemercier2022}. A deep filter can also be directly learnt, e.g. in \cite{Schroeter2020CLCNet} as a frequency-independent time filter or in \cite{Mack2020DeepFiltering, Schroeter2022DeepFilterNet} as a joint time-frequency filter.

In this paper, we propose a \ac{dsfe} scheme to transform masking-based speech restoration \acp{dnn} into deep subband filters. The proposed extension is implemented by using a learnable temporal convolution at the output of the original masking \ac{dnn} backbone and training the resulting architecture in an end-to-end fashion in the \ac{tf} domain.
Most of the time, the original masking \ac{dnn} already handles multi-frame filtering internally through e.g. temporal convolutions. However, we show that enforcing explicit multi-frame subband filtering as the final stage of processing results in a significant performance increase for dereverberation while leaving the denoising performance virtually unaltered. We 
justify our approach by relating time-frequency multiplicative masking and deep subband filtering to the noising and reverberation corruption models respectively.
The proposed approach has a negligible computational overhead and constitutes a generic module that can be plugged in any masking-based system.

The remainder of this paper is organized as follows. We first present an overview of the signal model and prerequisite assumptions for reverberation and noising corruptions. Then, we introduce our deep subband filtering extension scheme. We proceed with describing our experimental setup including data generation and training configuration. Finally we present and discuss our results.

\section{Signal model}
\label{sec:signal_model}

\subsection{Narrowband and subband filtering}

Filtering in the time-domain is obtained via convolution of a filter $\tim{w}$ with the speech signal $\tim{s}$, yielding the filtered signal $\tim{x}$:
\begin{equation} \label{eq:filter}
    \tim{x}_t \ssp
    = \ssp \sum_\tau \ssp \tim{w}_\tau \ssp \tim{s}_{t-\tau},
\end{equation}
where $t$ is the time index.
A well-known result of Fourier theory is that, when transposed in the Fourier domain, such a filtering process can be expressed as a multiplication of the Fourier spectra.
When using the \ac{stft} however, the window used for analysis is of limited size, and spectral leakage between frequency bands can occur. Consequently, the true filtering model is:
\begin{equation}
    \spec{x}_{t,f} \ssp = \ssp \sum_\tau \ssp \sum_\nu \ssp \tilde{\spec{w}}_{\tau, f, \nu} \ssp \spec{s}_{t - \tau, \nu} ,
\end{equation}
where $f$ is the frequency index, $\spec{x} \ssp := \ssp \mathrm{STFT}(\tim{x})$, $\spec{s} \ssp := \ssp \mathrm{STFT}(\tim{s})$ and $\tilde{\spec{w}}_{\tau, f, \nu}$ is interpreted as a response to a time-frequency impulse $\delta_{\tau, f - \nu}$ \cite{avargel_system_2007}. The sum over index $\nu$ represents cross-band filtering, and the sum over index $\tau$ is a convolution along the time dimension.

The \textit{subband approximation} ignores the effects of spectral leakage. Therefore, cross-band filtering is discarded and a single convolution is computed along the time-dimension in each frequency band independently:
\begin{equation} \label{eq:subband}
    \spec{x}_{t,f} \ssp
    = \ssp \sum_\tau \ssp \spec{w}_{\tau, f} \ssp \spec{s}_{t - \tau, f},
\end{equation}
where $\spec{w}:=\mathrm{STFT}(\tim{w})$.

The \textit{narrowband approximation} further assumes that the length of the filter $\tim{w}$ is inferior to the \ac{stft} window length, therefore zeroing out the filter taps $\{ \spec{w}_{\tau, f} ; \tau \geq 1 \}$ and yielding the following filtering model
:
\begin{equation} \label{eq:narrowband}
    \spec{x}_{t,f} \ssp = \ssp \spec{w}_{f} \ssp \spec{s}_{t,f}. 
\end{equation}

\subsection{Corruption models}

Speech denoising consists in removing additive background noise $n$ from the mixture $x$. The forward corruption process can naturally be represented in the \ac{stft} domain by addition of the clean speech and noise spectrograms: 
\begin{equation}
    \spec{x} \ssp = \ssp \spec{n} \ssp + \ssp \spec{s}.
\end{equation}

Many speech denoising approaches use time-frequency masking, i.e. they compute a mask $\spec{m}$ for each time-frequency bin and apply it to retrieve the clean speech estimate $\hat{\spec{s}}$:
\begin{equation} \label{eq:masking}
    \hat{\spec{s}}_{t,f} \ssp = \ssp (\spec{m} \ssp \odot \ssp \spec{x})_{t,f} \ssp := \ssp \spec{m}^t_f \ssp \spec{x}_{t,f}.
\end{equation}
This model is similar to the narrowband approximation \eqref{eq:narrowband} with a \textit{time-dependent} filter $\spec{m}$. We put the index $t$ as superscript to avoid confusion with the time-convolution index $\tau$.

In contrast to denoising, speech dereverberation aims to recover the anechoic speech corrupted by room acoustics. The signal model is exactly the filtering process in \eqref{eq:filter}, where the filter $\tim{w}$ is called the \ac{rir}. Since the \ac{rir} length is almost always  larger than the \ac{stft} window length, one cannot use the narrowband approximation \eqref{eq:narrowband} and has to resort to the subband approximation \eqref{eq:subband} instead.
Consequently, some speech dereverberation methods perform inverse filtering in the \ac{stft} domain using the subband approximation \cite{Schroeter2022DeepFilterNet,Nakatani2008b,Kinoshita2017,Heymann2018,Lemercier2022}. That is, they try and estimate a filter $\mf{m}$ supposed to represent the inverse of the \ac{rir}, such that the anechoic speech estimate is retrieved as:
\begin{equation} \label{eq:convolutive_masking}
    \hat{\spec{s}}_{t,f} \ssp = \ssp (\mf{m} \ssp \ast \ssp \spec{x})_{t,f} \ssp := \ssp \sum_\tau \ssp \mf{m}^t_{\tau,f} \ssp \spec{x}_{t - \tau,f},
\end{equation}
with $\ast$ representing a convolution over the time-axis. Please note that in the model above in contrast to \eqref{eq:subband}, the filter $\mf{m}$ is considered \textit{time-dependent}, same as in the time-frequency masking case. This is often assumed in order to account for non-stationarity of the \ac{rir} and estimation errors \cite{Jukic2017Adaptive, Heymann2018, Lemercier2022, Lemercier2022b}.

\section{Deep subband filtering extension} 
\label{sec:mf}

\begin{figure}
\centering
\scalebox{0.8}{
\begin{tikzpicture}[
BOX/.style={rectangle, draw=black!100, minimum size=10mm},
NEWBOX/.style={rectangle, draw=black!100, fill=red!20, minimum size=10mm},
LEARNEDBOX/.style={rectangle, rounded corners, draw=black!100, fill=blue!20, minimum size=10mm},
CIRC/.style={circle, draw=black!100, minimum size=5mm},
align=center,node distance=0.45cm
]

\node (encoded) [] {$\spec{x}_{t,f}$};

\node[LEARNEDBOX] (TCN) [right=of encoded, align=center] {Masking DNN \\ $f_\theta$};
\draw[->, thick] (encoded.east) to (TCN.west);

\node (filter) [right=of TCN] {$\spec{m}^t_f$};
\draw[->, thick] (TCN.east) to (filter.west);

\node[LEARNEDBOX] (mffe) [right=of filter, align=center] {Deep Subband \\ Filtering Extension \\ $g_\phi$};
\draw[->, thick] (filter.east) to (mffe.west);

\node (mffilter) [right=of mffe] {$\mf{m}^t_{\tau,f}$};
\draw[->, thick] (mffe.east) to (mffilter.west);

\node[CIRC] (filtering) [below=of mffilter] {$\ast_\tau$};
\draw[->, thick] (mffilter) to (filtering);

\draw[->, thick] (encoded.south) |- (filtering);

\node (filtered) [right=of filtering] {$\hat{\spec{s}}_{t,f}$};
\draw[->, thick] (filtering.east) to (filtered.west);

\end{tikzpicture}
}
\caption{\protect\centering{Proposed model diagram. The blue blocks are learnable neural networks.}}
\label{fig:diagram}
\end{figure}
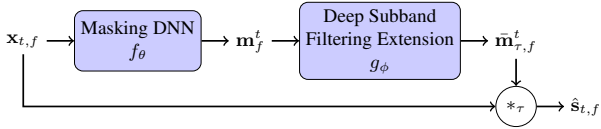

Many neural network-based schemes use time-frequency masking, without examining the nature of the corruption. In this section, we present our \ac{dsfe} scheme, which turns the time-frequency masks produced by such \acp{dnn} into subband filters. 
Let $f_\theta$ be a \ac{dnn} providing a mask $\spec{m} = f_\theta(\spec{x})$ in the complex spectrogram domain, such that the clean spectrogram estimate is obtained via time-frequency masking \eqref{eq:masking}.

We wish to extend the mask $\spec{m}$ into a filter $\mf{m}$ implemented by the neural network combination $\mf{m} = g_\phi(f_\theta(\spec{x}))$, such that the clean estimate is obtained via subband filtering \eqref{eq:convolutive_masking}. Essentially, we want to turn masking \acp{dnn} into deep subband filters \cite{Mack2020DeepFiltering}.
To this end, we design the \textit{deep subband filtering extension} $g_\phi$ as a point-wise two-dimensional convolutional layer with tanh activation. The maps are of size $T \times F$, the kernels of size $1 \times 1$ and there are $2$ input and $2 N_f$ output channels corresponding to the single-frame mask and multi-frame filter real and imaginary parts, respectively:
\begin{equation}
    g_\phi \ssp : \ssp \spec{m} \ssp \rightarrow \ssp \mf{m}  \ssp = \ssp \frac{1}{N_f} \tanh \ssp \left( \mathrm{Conv2D}( \spec{m} \ssp; \ssp \phi ) \right).
\end{equation}

Note that we feed the spectrogram $\spec{x}$ to the neural network and only use the multi-frame representation $\{ \spec{x}_{t - \tau, f} ; \tau \in [0, 1, ..., N_f -1] \}$ for filtering. This is because the multi-frame representation does not add any relevant information with respect to $\spec{x}$: since most \acp{dnn} compute correlations along the time-dimension already, it is redundant to provide a vector which explicitly encodes that time-delayed information. 
The proposed algorithm is summarized on Figure~\ref{fig:diagram}.

As the corresponding inverse filtering model better fits the corruption model for reverberation, we expect our \ac{dsfe} method to perform better at dereverberation than its %
masking counterpart, and not 
produce significant changes for denoising.

\section{Experimental setup}
\label{sec:exp}

\subsection{Data}

Both datasets for denoising and dereverberation experiments use the 
WSJ0
corpus \cite{datasetWSJ0} for clean speech sources. The training, validation and test splits comprise 101, 10 and 8 speakers for a total of 12777, 1206 and 651 utterances and a length of 25, 2.3 and 1.5 hours of speech respectively, sampled at $16$ kHz.

\textit{Speech Denoising}: \,The WSJ0+Chime dataset is generated using clean speech extracts from the WSJ0 corpus and noise signals from the CHiME3 dataset \cite{barker2015third}. The mixture signal is created by randomly selecting a noise file and adding it to a clean utterance with a \ac{snr} sampled uniformly between -6 and 14$\,$dB.

\textit{Speech Dereverberation}: \,The WSJ0+Reverb dataset is generated using clean speech data from the WSJ0 corpus and convolving each utterance with a simulated \ac{rir}. We use the \texttt{\small pyroomacoustics} library \cite{Scheibler2018Pyroom} to simulate the \acp{rir}. The reverberant room is modeled by sampling uniformly a target $T_\mathrm{60}$ between 0.4 and 1.0 seconds and room length, width and height in [5,15]$\times$[5,15]$\times$[2,6] m. 
The anechoic target is generated using the $T_\mathrm{60}$-shortening method \cite{Zhou2022T60}, where the \ac{rir} is shaped by a decaying exponential window so that the resulting $T_\mathrm{60}$ equals $200$ms. This results in an average \ac{drr} of -5.3$\,$dB.

\subsection{Single-frame DNN backbone}

In this paper, we use the GaGNet architecture by \cite{Li2022Gagnet}, a state-of-the-art denoising neural network, which is the successor of \cite{Li2021DNS} which ranked first in the real-time enhancement track of the DNS-2021 challenge. 
GaGNet leverages magnitude-only and complex-domain information in parallel with temporal convolutional networks. The rationale is to obtain a coarse estimation with the magnitude-processing \textit{glance} modules, and to refine this estimation with \textit{gaze} modules processing the real and imaginary parts of the complex spectrogram. Between each repeated glance and gaze module, an approximate complex ratio mask \cite{Williamson2017j} is applied on the current version of the signal to enforce a coherent filtering process and stabilize training. Finally, the network outputs multiplicative mask values for the real and imaginary parts.
We name our proposed method \textit{DSFE-GaGNet}, which is the concatenation of GaGNet with the DSFE module $g_\phi$.
Although we focus on GaGNet in this work, please note that our \ac{dsfe} method is compatible with any architecture performing mask estimation in the complex \ac{stft} domain. It could even be envisaged to use a similar extension in a different domain e.g. learnt by an \ac{dnn} encoder.

\subsection{Training configuration}

We use the same training configuration as GaGNet \cite{Li2022Gagnet}%
: the \ac{stft} uses a Hann window with $320$ points and $50\%$ overlap at a sample rate of $16$ kHz. We employ square-root compression on the magnitude spectrogram. Therefore, the features that are fed to GaGNet are: $\mathrm{cat} ( \sqrt{|\spec{x}|} \ssp \cos (\phi_x) \ssp , \ssp \sqrt{|\spec{x}|} \ssp \sin (\phi_x) )$, where $\spec{x} \ssp = \ssp |\spec{x}| \ssp \exp(j \phi_x)$ is the noisy complex spectrogram.
The training loss is a sum of mean square errors with respect to the real part, imaginary part and magnitude of the clean and estimated spectrograms.
The networks are trained with the Adam optimizer with a learning rate of $0.0005$. Contrarily to \cite{Li2022Gagnet}, we use mini-batches of size $48$ and use early stopping with a patience of $50$ epochs and a maximum of $2000$ epochs.

\newcommand{\spectrow}{.65\columnwidth}
\newcommand{\spectroxs}{.05\columnwidth}
\newcommand{\spectroys}{0.57\columnwidth}

\begin{figure}[H]
\centering 

\begin{tikzpicture}[scale=0.77, transform shape]
 \centering
\begin{axis}
[    
    axis line style={draw=none},
    name={clean},
    title = {Clean},
    xmin = 0, xmax = 2,
    ymin = 0, ymax = 8000,
    xticklabels=\empty,
    ytick = {0,2000,...,8000},
    ylabel = {Frequency [Hz]},
    width =\spectrow,
    height =\spectrow
]
\addplot graphics[xmin=0,ymin=0,xmax=2,ymax=8000] {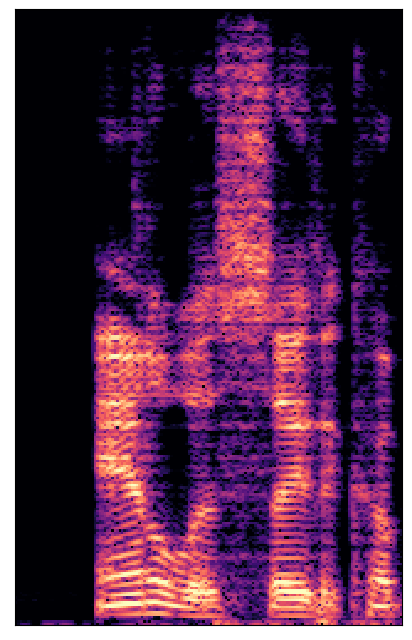};

\draw (axis cs:0.8,750) circle (0.8cm and 0.25cm) [very thick, red];

\end{axis}
 
 \begin{axis}
[    axis line style={draw=none},
    at={(clean.south east)},
    name={reverberant},
    title = {Reverberant},
    xmin = 0, xmax = 2,
    ymin = 0, ymax = 8000,
    xtick = {},
    yticklabels=\empty,
    xticklabels=\empty,
    width =\spectrow,
    height =\spectrow,
    xshift = \spectroxs
]
\addplot graphics[xmin=0,ymin=0,xmax=2,ymax=8000] {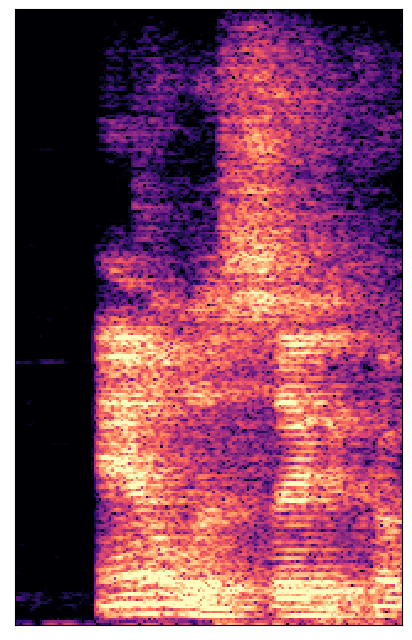};
\end{axis}
  
 \begin{axis}
[    axis line style={draw=none},
    at={(reverberant.south east)},
    xshift = 0.005\textwidth,
    yshift = -0.18\textwidth,
    width = 0.15\textwidth,
    height = 0.4\textwidth,
    hide axis,
]
\addplot graphics[xmin=0,ymin=0,xmax=1,ymax=1] {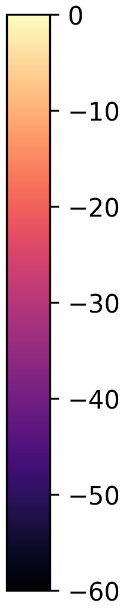};
\end{axis}

 \begin{axis}
[    axis line style={draw=none},
    at={(clean.south west)},
    yshift=-\spectroys,
    name={sf},
    title = {DSFE-GaGNet $N_f=20$},
    xmin = 0, xmax = 2,
    ymin = 0, ymax = 8000,
    xtick = {0,1, ...,2},
    ytick = {0,2000,...,8000},
    xlabel = {Time [s]},
    ylabel = {Frequency [Hz]},
    width =\spectrow,
    height =\spectrow
]
\addplot graphics[xmin=0,ymin=0,xmax=2,ymax=8000] {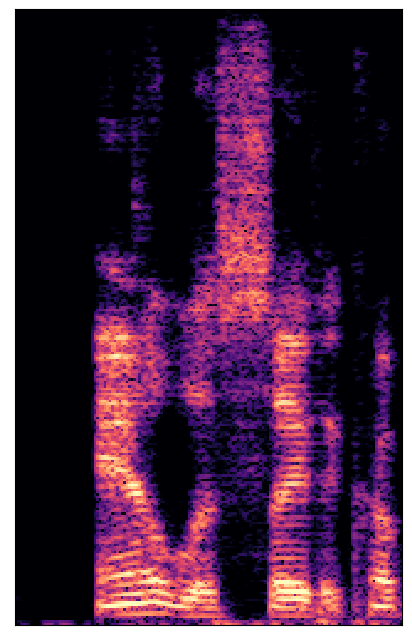};

\draw (axis cs:0.8,750) circle (0.8cm and 0.25cm) [very thick, red];

\end{axis}

  \begin{axis}
[    axis line style={draw=none},
    at={(sf.south east)},
    name={mf},
    title = {GaGNet},
    xmin = 0, xmax = 2,
    ymin = 0, ymax = 8000,
    yticklabels=\empty,
    xtick = {0,1, ...,2},
    xlabel = {Time [s]},
    width =\spectrow,
    height =\spectrow,
    xshift = \spectroxs
]
\addplot graphics[xmin=0,ymin=0,xmax=2,ymax=8000] {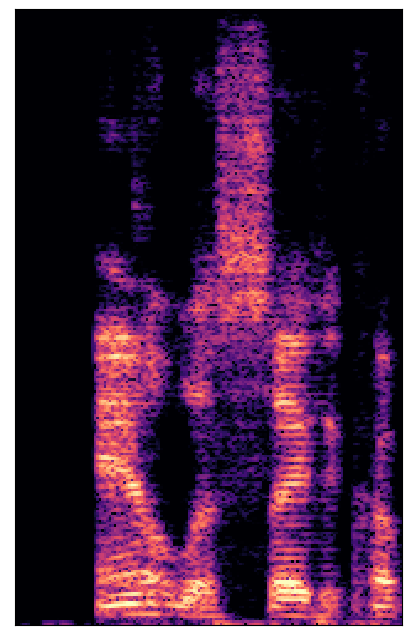};

\draw (axis cs:0.8,780) circle (0.8cm and 0.25cm) [very thick, red];

\end{axis}

\end{tikzpicture}
 
\caption{\protect\centering Log-energy spectrograms of clean, reverberant and processed signals form the WSJ0+Reverb dataset. The harmonic structure in the red circle is altered with GaGNet and better preserved with DSFE-GaGNet. $\mathrm{T}_\mathrm{60} = 0.85\mathrm{s}$.}
\label{fig:spectro}
\end{figure}
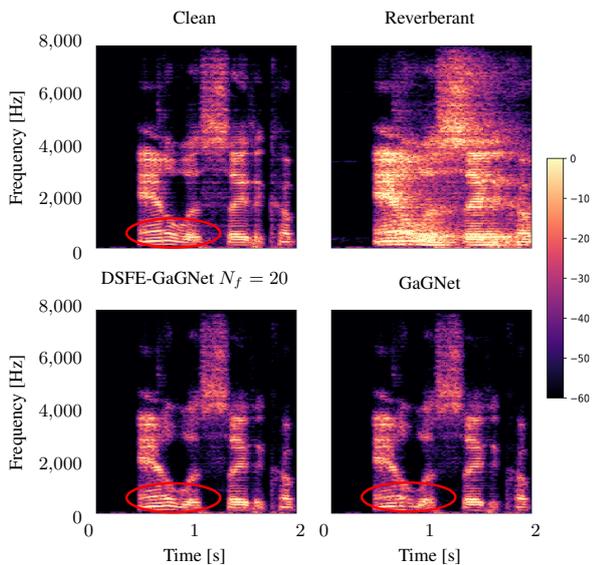

\vspace{-2em}

\subsection{Evaluation}

We conduct instrumental evaluation using classical speech metrics like \ac{polqa} \cite{POLQA2018}, \ac{pesq} \cite{Rix2001PESQ}, \ac{estoi} \cite{Jensen2016ESTOI} as well as \ac{si} \ac{sdr}, \ac{sir} and \ac{sar} \cite{Leroux2019SISDR}. We also report the number of million single-point floating operations per second of processed speech (MFLOPS$\cdot \mathrm{s}^{-1}$) as provided by the \texttt{pypapi} library\footnote{https://github.com/flozz/pypapi}.

\section{Experimental results and discussion}
\label{sec:results}

\begin{table*}[t]
    \centering
    \caption{\protect\centering{Dereverberation results obtained on the WSJ0-Reverb dataset. Values indicate mean and standard deviation.}}
    \begin{tabular}{c|c|cccccc|c}

\toprule 
Method & $N_f$ & POLQA & PESQ & ESTOI & SI-SDR & SI-SIR & SI-SAR & MFLOPS$\cdot \mathrm{s}^{-1}$ \\

\midrule
\midrule
Mixture & $-$ & 1.94 $\pm$ 0.40 & 1.51 $\pm$ 0.30 & 0.62 $\pm$ 0.12 & 1.2 $\pm$ 2.8 & -0.8 $\pm$ 2.5 & $-$ & $-$\\
\midrule

GaGNet & 1 & 
3.07 $\pm$ 0.43 & 2.52 $\pm$ 0.44 & 0.83 $\pm$ 0.06 & 6.0 $\pm$ 2.4 & 5.9 $\pm$ 2.6 & 6.0 $\pm$ 2.4 & 367.2 \\

\midrule
DSFE-GaGNet & 4 & 
3.17 $\pm$ 0.41 & 2.60 $\pm$ 0.44 & 0.85 $\pm$ 0.05 & 6.9 $\pm$ 2.3 & 6.8 $\pm$ 2.7 & 6.4 $\pm$ 2.2 & 368.6 \\
DSFE-GaGNet & 8 & 
3.30 $\pm$ 0.42 & 2.77 $\pm$ 0.44 & 0.86 $\pm$ 0.05 & 7.5 $\pm$ 2.1 & 7.4 $\pm$ 2.7 & 6.7 $\pm$ 2.1 & 368.8 \\
DSFE-GaGNet & 12 & 
3.29 $\pm$ 0.42 & 2.75 $\pm$ 0.44 & 0.86 $\pm$ 0.05 & 7.3 $\pm$ 2.2 & 7.1 $\pm$ 2.7 & 6.7 $\pm$ 2.2 & 370.0 \\
DSFE-GaGNet & 16 & 
3.36 $\pm$ 0.42 & 2.81 $\pm$ 0.45 & 0.87 $\pm$ 0.05 & 7.6 $\pm$ 2.2 & 7.6 $\pm$ 2.7 & 6.7 $\pm$ 2.2 & 370.3 \\
DSFE-GaGNet & 20 & 
\textbf{3.41 $\pm$ 0.40} & \textbf{2.85 $\pm$ 0.44} & \textbf{0.87 $\pm$ 0.05} & \textbf{7.9 $\pm$ 2.3} & \textbf{7.9 $\pm$ 2.9} & \textbf{6.9 $\pm$ 2.2} & 371.6 \\
DSFE-GaGNet & 24 & 
3.17 $\pm$ 0.43 & 2.61 $\pm$ 0.45 & 0.84 $\pm$ 0.06 & 6.6 $\pm$ 2.4 & 6.5 $\pm$ 2.7 & 6.2 $\pm$ 2.3 & 371.8\\
 \midrule
    \bottomrule
    \end{tabular}
    \label{tab:results:derev}
\end{table*}

\begin{table*}[t]
    \centering
    \caption{\protect\centering{Denoising results obtained on the WSJ0+Chime dataset. Values indicate mean and standard deviation.}}
    \begin{tabular}{c|c|cccccc|c}

\toprule 
Method & $N_f$ & POLQA & PESQ & ESTOI & SI-SDR & SI-SIR & SI-SAR & MFLOPS$\cdot \mathrm{s}^{-1}$ \\
    
    \midrule
    \midrule
    Mixture & $-$ & 2.08 $\pm$ 0.64 & 1.38 $\pm$ 0.32 & 0.65 $\pm$ 0.18 & 4.3 $\pm$ 5.8 & 4.3 $\pm$ 5.8 & $-$ & $-$ \\
    \midrule

    GaGNet & 1 & 
    3.48 $\pm$ 0.60 & 2.75 $\pm$ 0.59 & \textbf{0.89 $\pm$ 0.08} & \textbf{15.5 $\pm$ 4.1} & 26.1 $\pm$ 4.5 & \textbf{16.0 $\pm$ 4.3} & 367.2 \\

    \midrule
    DSFE-GaGNet & 4 &
    3.33 $\pm$ 0.64 & 2.69 $\pm$ 0.59 & 0.88 $\pm$ 0.08 & 14.4 $\pm$ 4.1 & 25.2 $\pm$ 5.0 & 14.8 $\pm$ 4.2 & 368.6 \\
    DSFE-GaGNet & 8 &
    3.42 $\pm$ 0.63 & 2.72 $\pm$ 0.60 & 0.88 $\pm$ 0.08 & 14.9 $\pm$ 4.1 & 26.2 $\pm$ 4.8 & 15.4 $\pm$ 4.2 & 368.8 \\
    
    DSFE-GaGNet & 12 &
    3.46 $\pm$ 0.61 & 2.75 $\pm$ 0.58 & 0.89 $\pm$ 0.08 & 15.0 $\pm$ 4.0 & 27.0 $\pm$ 5.1 & 15.4 $\pm$ 4.0 & 370.0 \\
    DSFE-GaGNet & 16 &
    3.44 $\pm$ 0.63 & 2.72 $\pm$ 0.59 & 0.89 $\pm$ 0.08 & 15.3 $\pm$ 4.2 & 26.7 $\pm$ 4.8 & 15.7 $\pm$ 4.3 & 370.3 \\
    DSFE-GaGNet & 20 &
    3.30 $\pm$ 0.63 & 2.65 $\pm$ 0.57 & 0.88 $\pm$ 0.08 & 14.1 $\pm$ 3.9 & 24.7 $\pm$ 5.0 & 14.6 $\pm$ 3.9 & 371.6 \\
    DSFE-GaGNet & 24 &
    \textbf{3.51 $\pm$ 0.60} & \textbf{2.82 $\pm$ 0.56} & \textbf{0.89 $\pm$ 0.08} & \textbf{15.5 $\pm$ 4.1} & \textbf{27.1 $\pm$ 4.8} & \textbf{16.0 $\pm$ 4.2} & 371.8 \\

    \midrule
    \bottomrule
\end{tabular}
\label{tab:results:denoising}
\end{table*}
\newcommand{\w}{0.28\textwidth}
\newcommand{\h}{0.28\textwidth}
\newcommand{\xs}{0.05\textwidth}
\newcommand{\ys}{0.02\textwidth}
\newcommand{\lxs}{0.12\textwidth}
\newcommand{\lys}{-0.035\textwidth}
\newcommand{\tys}{87pt}
\newcommand{\txs}{30pt}

\definecolor{orange}{HTML}{feb24c}
\definecolor{red}{HTML}{f03b20}

\begin{figure}
\hspace{-1em}
\begin{tikzpicture}

\begin{axis}[title={$\Delta$POLQA}, name=polqa,
ymajorgrids,
width=\w, height=\h, xshift=\xs,
xtick={1,4,8,12,16,20,24}
]

\addplot+[orange, mark=*, mark options={fill=orange, draw=orange}] table [x=N, y=POLQA, col sep=comma] {derev.csv};
\addplot+[red, mark=*] table [x=N, y=POLQA, col sep=comma] {enh.csv};

\end{axis}

\begin{axis}[title={$\Delta$PESQ}, name=pesq, at={(polqa.south east)},
ymajorgrids,
width=\w, height=\h, xshift=\xs,
xtick={1,4,8,12,16,20,24}
]

\addplot+[orange, mark=*, mark options={fill=orange, draw=orange}] table [x=N, y=PESQ, col sep=comma] {derev.csv};
\addplot+[red, mark=*] table [x=N, y=PESQ, col sep=comma] {enh.csv};

\end{axis}

\begin{axis}[title={$\Delta$ESTOI}, name=estoi, at={(polqa.south west)},
ymajorgrids,
width=\w, height=\h, xshift=0, yshift=-.26\textwidth,
xtick={1,4,8,12,16,20,24},
xlabel={$N_f$},
legend style={
    at={(xticklabel cs:.5)},
    anchor=north,
    xshift=\lxs,
    yshift=\lys
},
legend columns=2
]

\addplot+[orange, mark=*, mark options={fill=orange, draw=orange}] table [x=N, y=ESTOI, col sep=comma] {derev.csv};
\addplot+[red, mark=*] table [x=N, y=ESTOI, col sep=comma] {enh.csv};

\legend{Dereverberation, Denoising}
\end{axis}

\begin{axis}[title={$\Delta$SI-SDR}, name=sisdr, at={(estoi.south east)},
ymajorgrids,
width=\w, height=\h, xshift=\xs,
xtick={1,4,8,12,16,20,24},
xlabel={$N_f$}
]

\addplot+[orange, mark=*, mark options={fill=orange, draw=orange}] table [x=N, y=SISDR, col sep=comma] {derev.csv};
\addplot+[red, mark=*] table [x=N, y=SISDR, col sep=comma] {enh.csv};

\end{axis}

\end{tikzpicture}
    
    \caption{
    \protect\centering \textit{Instrumental metrics improvements of DSFE-GaGNet with respect to single-frame GaGNet for speech denoising on WSJ0+Chime and dereverberation as a function of the number of frames $N_f$.
    }}
    \label{fig:frames}
\end{figure}
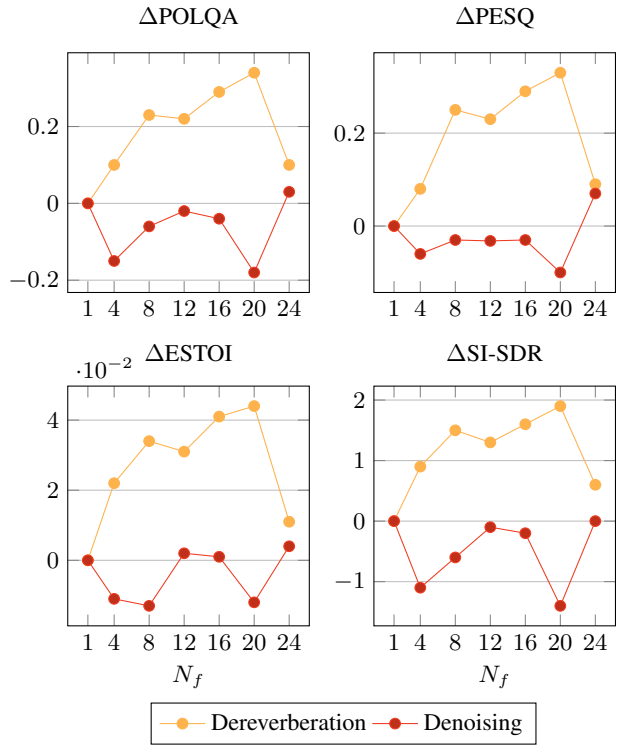
\subsection{Multi-frame filtering for speech enhancement tasks}

We report results for dereverberation on WSJ0+Reverb and denoising on WSJ0+Chime in tables \ref{tab:results:derev} and \ref{tab:results:denoising} respectively.
For a more direct comparison, we group these experiments in Figure~\ref{fig:frames} by showing the improvements of our method DSFE-GaGNet, with respect to its single-frame GaGNet counterpart as a function of $N_f$ for both dereverberation and denoising. 

For dereverberation, we observe a monotonic increase in all instrumental metrics as more frames are used in DSFE-GaGNet. The performance peaks at $N_f=20$ with an improvement of $.33$ PESQ, $.04$ ESTOI and $1.9\mathrm {dB}$ SI-SDR over the single-frame baseline. This improvement then decreases, as we observe that training is less stable with a high number of frames e.g. $N_f=24$.
We observe on the spectrograms displayed in \figurename~\ref{fig:spectro} that DSFE-GaGNet preserves the harmonic structure in some cases where that structure is altered by GaGNet.

In the denoising case, the DSFE module reveals useless as DSFE-GaGNet performance saturates at the level of the single-frame GaGNet, or even worsens with more frames, at the exception of $N_f=24$ where marginal improvements are observed.

This comparison suggests that subband filtering should be adopted when it fits the corruption model, i.e. for convolutive signal models like reverberation where the narrowband approximation requirements are not satisfied. In that case, we can obtain remarkable improvements at a very low computational cost: our best model DSFE-GaGNet with $N_f=20$ only requires $4.4$ MFLOPS$\cdot \mathrm{s}^{-1}$ more than GaGNet, that is, a relative $1.2\%$ increase. Furthermore, the temporal convolution used in the DSFE module with $N_f=20$ frames employs only $96$ trainable parameters, which is negligible compared to the $5.9$M parameters of the original GaGNet backbone. 
Finally, since DSFE-GaGNet only uses past frames, the algorithmic latency does not increase and is still dominated by the length of the STFT synthesis window i.e. 20ms.

\subsection{Ablation study}

\begin{table}[t]
    \caption{\protect\centering{\textit{Dereverberation results of DSFE-GaGNet on WSJ0+Reverb. All approaches use $N_f=20$ frames. Values indicate mean and standard deviation}.}}
    \centering
    \scalebox{0.9}{
    \begin{tabular}{c|ccc}
    \toprule
        Strategy & POLQA & ESTOI & SI-SDR \\
        \midrule
        \midrule
        Mixture & 1.94 $\pm$ 0.40 & 0.62 $\pm$ 0.12 & 1.2 $\pm$ 2.8 \\
        \midrule
        Pretrain$+$Freeze & 3.19 $\pm$ 0.42 & 0.84 $\pm$ 0.06 & 6.8 $\pm$ 2.4\\
        Pretrain$+$Finetune & 3.40 $\pm$ 0.41 & 0.86 $\pm$ 0.05 & 7.5 $\pm$ 2.5 \\
        Join & \textbf{3.41 $\pm$ 0.40} & \textbf{0.87 $\pm$ 0.05} & \textbf{7.9 $\pm$ 2.3} \\
        \midrule \bottomrule
    \end{tabular}}
        \vspace{-1em}
    \label{tab:training}
\end{table}

In Table \ref{tab:training} we present results of an ablation study showing various training strategies for DSFE-GaGNet. The default training configuration is denoted as \textit{Join}, i.e. when both the \ac{dsfe} module and the GaGNet backbone are trained jointly from scratch. We also try pretraining the GaGNet backbone and subsequently tuning the DSFE module parameters, either leaving the GaGNet backbone frozen (\textit{Pretrain$+$Freeze}) or finetuning it along the \ac{dsfe} parameters (\textit{Pretrain$+$Finetune}).
As expected, joint training performs best, but the improvement over \textit{Pretrain$+$Finetune} is marginal.
This highlights that it is paramount to jointly tune the  DSFE parameters along with the single-frame backbone, at least at some stage of the training.

\section{Conclusion}
\label{sec:conclusion}
We present a deep subband filtering extension scheme transforming \acp{dnn} performing time-frequency multiplicative masking into deep subband filters. We show that such an extension fits the subband filtering approximation used for dereverberation in the \ac{stft} domain, while time-frequency masking fits the narrowband filtering approximation used for denoising.
Consequently, we show that our deep subband filtering extension significantly increases dereverberation performance while leaving denoising performance virtually the same. 
The proposed extension scheme can be generically applied to any \ac{dnn} baseline performing time-frequency masking, with an insignificant increase in inference time and model capacity.
Ablation studies suggest that the deep subband filtering extension module should be trained jointly with the original single-frame \ac{dnn}, at least at some stage of the training.

\vfill
\newpage

\bibliographystyle{ieeetr}
\bibliography{biblio}

\end{document}